\documentclass[%
 reprint,
 amsmath,amssymb,
 aps,
]{revtex4-2}

\usepackage{appendix}
\usepackage{xcolor}
\usepackage{graphicx}
\usepackage{dcolumn}
\usepackage{bm}

\usepackage{multirow}
\usepackage{amsmath}

\begin{document}

\preprint{APS/123-QED}

\title{Misalignment production of isotropized vector dark matter?}%

\author{Chong-Bin Chen$^{1,2}$}
 \email{chongbin@ncu.edu.cn}
\affiliation{%
$^{1}$ Department of Physics, Nanchang University, Nanchang, 330031, China\\
$^{2}$ Center for Relativistic Astrophysics and High Energy Physics,
Nanchang University, Nanchang 330031, China
}%


\begin{abstract}
We present dark matter production by the misalignment mechanism of a multi-vector condensate through kinetic coupling during inflation. We impose isotropized background vector fields to release the model
from the stringent constraint of anisotropy. However, it turns out that the constraints imposed by non-Gaussianity and isocurvature fluctuations are incompatible with each other, regardless of whether the fluctuations are in the weak-mixing or strong-mixing regime.

\end{abstract}

\maketitle


\section{Introduction}
The nature of dark matter (DM) remains one of the most puzzling questions in cosmology. Currently, there exist numerous beyond‑Standard‑Model bosons that serve as DM candidates, such as vector bosons produced during inflation, also referred to as dark photons. These vector bosons typically arise from a hidden gauge sector, such as U(1), and can yield the correct relic abundance of DM today.

However, the transverse modes of vector fields are generally conformally coupled to gravity, so that their energy density is rapidly diluted during inflation. There are many mechanisms to break this situation and thereby enhance DM production: gravitational production by a minimal realization of the longitudinal mode of massive vector fields \cite{Graham:2015rva}, or non-minimal coupling with gravity \cite{Ozsoy:2023gnl}; tachyonic production at the end of inflation from a rolling axion field \cite{Bastero-Gil:2018uel,Bastero-Gil:2021wsf}; resonance production of oscillation of fields \cite{Co:2018lka,Dror:2018pdh,Zhang:2025pgb,Adshead:2023qiw}; production from cosmic string \cite{Long:2019lwl,Kitajima:2022lre}; kinetic coupling \cite{Nakai:2020cfw,Firouzjahi:2020whk,Salehian:2020asa,Nakai:2022dni}; misalignment production \cite{Nelson:2011sf,Nakayama:2019rhg,Nakayama:2020rka,Kitajima:2023fun,Kaneta:2023lki,Fujita:2023axo}.

The kinetic coupling introduces $f^2(\phi)F^2$ to break the conformal coupling without introducing ghost instabilities. Here $f(\phi)$ is a kinetic function of a scalar field $\phi$. For a minimal realization, $\phi$ can be identified with the inflaton \cite{Nakai:2020cfw}. In this case, if $f\propto a^{-n}$ during inflation, $n$ is restricted to a very narrow allowed window $-2.1<n<-2$, to obtain the correct DM abundance and avoid backreaction. In the charged inflation case, the parameter space is restricted to an even narrower band \cite{Firouzjahi:2020whk}. 

Backreaction does not necessarily spoil inflation. On the contrary, inflation can settle into an anisotropic attractor, in which the vector energy density tracks that of the inflaton sector \cite{Watanabe:2009ct, Watanabe:2010fh, Kanno:2010nr, Yamamoto:2012tq, Maleknejad:2012fw,Chen:2021nkf,Chen:2022ccf}. Such a homogeneous vector condensate could be a realization of the misalignment production of vector DM if it has mass, just like the axion. The first model of vector misalignment production via the Steuckelberg mechanism is proposed in \cite{Nelson:2011sf}. However, ghost instability makes this model unviable \cite{Nakayama:2019rhg}. Therefore, \cite{Nakayama:2019rhg,Nakayama:2020rka,Kitajima:2023fun} study the possibility of misalignment production in the anisotropic attractor of kinetic coupling $f^2(\phi)F^2$. Nevertheless, this model is severely ruled out by large-scale anisotropy, unless the curvaton mechanism is invoked \cite{Kitajima:2023fun}. In \cite{Fujita:2023axo}, a model that circumvents the anisotropy problem by a SU(2) gauge field was proposed, albeit through an axial $\phi F\tilde{F}$ coupling.

In this paper, we explore isotropized multi-vector condensate to realize vector DM via kinetic coupling $f^2(\phi)F^2$. It has been shown that the isotropic configuration is an attractor in the phase space for the number of vector fields greater than two \cite{Yamamoto:2012tq}. Hence, we will consider an isotropized vector field to evade the constraints from the stringent constraint of anisotropy. We will investigate its viability as a mechanism for DM production. In particular, we will discuss the constraints on the model in the strong-mixing regime of perturbations, which was recently studied in \cite{Chen:2022ccf,Chen:2023bcz,Chen:2026rwr,Chen:2025qyv}.

The organization of the paper is as follows. In section \ref{section2} we discuss the dynamics of ``anisotropic'' attractor of the isotropized vector field with kinetic coupling. In section \ref{section3} we study how to obtain the correct relic abundance of DM through misalignment of vector fields. In section \ref{section4} we employ observational constraints to restrict its parameter space, primarily the power spectrum and non-Gaussianities of curvature perturbations, together with isocurvature perturbations. Section \ref{section5} is devoted to the conclusions.

\section{Isotropized massive vector with kinetic coupling}\label{section2}
We consider the following model for an inflaton and a number of massive vector boson 
\begin{align}
    S=\int dtd^3x{}&\sqrt{-g}\bigg[-\frac{1}{2}\partial_{\mu}\phi\partial^{\mu}\phi-V(\phi)\nonumber\\
    &-\frac{f^2(\phi)}{4}F_{\mu\nu}^{(a)}F^{(a)\mu\nu}-\frac{m_A^2}{2}A_{\mu}^{(a)}A^{(a)\mu}\bigg],
\end{align}
where $a$ labels the vector fields with mass $m_A$, and $F^{(a)}_{\mu\nu}\equiv\partial_{\mu}A^{(a)}_{\nu}-\partial_{\nu}A^{(a)}_{\mu}$ is the field strength tensor of the vector field $A^{(a)}_{\mu}$. If one considers a large number of randomly oriented vector fields, the vector sector is dominated by the $\delta_{ij}$ components \cite{Bento:1992wy,Golovnev:2008cf}. Therefore, in an FRW spacetime $ds^2=-dt^2+a^2(t)d\boldsymbol{x}^2$, we consider the following isotropized background configuration of the vector fields \cite{Maleknejad:2011sq,Maleknejad:2011jw}
\begin{align}\label{isoA}
    A^{(a)}_0=0,\ \ \ \ \ \ \ A^{(a)}_{i}=A(t)\delta_{ai}.
\end{align}
The kinetic function $f(\phi)$ breaks the conformal invariance of the vector field, allowing vector bosons to be produced during inflation. A commonly adopted form of the kinetic coupling is \cite{Watanabe:2009ct}
\begin{align}\label{f}
    f(\phi)=\exp\left(\frac{2c}{M_{\text{pl}}^2}\int d\phi{}\frac{V}{V_{\phi}}\right),
\end{align}
where $c$ is a dimensionless parameter and $V_{\phi}\equiv\partial_{\phi}V$. This kinetic coupling ensures the correct sign of the kinetic term, thereby preventing ghost instabilities \cite{Nakayama:2019rhg}.

The energy-momentum tensor of the vector fields is
\begin{align}
    T_{\mu\nu}
={}&f^2\left[
F^{(a)}_{\mu\rho}F^{(a)\rho}_{\nu}-\frac{1}{4}g_{\mu\nu}F^{(a)}_{\rho\sigma}F^{(a)\rho\sigma}\right]\nonumber\\
&+m_A^2\left[A^{(a)}_\mu A^{(a)}_\nu-\frac{1}{2}g_{\mu\nu}A^{(a)}_\rho A^{(a)\rho}\right].
\end{align}
For the isotropized background configuration in Eq.~\eqref{isoA}, the energy density and pressure are given by $\rho_A=-T^{0}_{\ 0}$ and $p_A=T^{i}_{\ i}/3$, respectively. Explicitly,
\begin{align}
    \rho_A={}&\frac{3}{2a^2}
\left(f^2\dot{A}^{\,2}+m_A^2 A^2\right),\\
p_A={}&\frac{1}{2a^2}
\left(f^2\dot{A}^{\,2}-m_A^2 A^2\right).
\end{align}
The first terms, $f^2\dot{A}^{\,2}$, in these expressions represent the kinetic energy, while the second terms, $m_A^2A^2$, correspond to the potential energy. Since the characteristic frequency of the vector field during inflation is $\omega\sim H$, the kinetic term is typically of order $f^2\omega^2A^2$. If the potential term is negligible compared with the kinetic term, namely when the effective mass of the vector bosons satisfies $m_A/f\ll H$, it can be safely neglected during and after inflation. 

In the following sections, we derive the dynamics of the vector fields and the inflaton during and after inflation, respectively. We do not impose any restriction on the ratio $h$, which depends on the model parameters (e.g., the parameter $c$ in the kinetic coupling, Eq.~\eqref{f}).

\subsection{Dynamics during inflation}\label{Dinf}
We can derive the equations of motion for the matter fields from the above action under the isotropized configuration of the vector fields:
\begin{align}
    \ddot{\phi}+3H\dot{\phi}+V_{\phi}-\frac{3}{a^2}ff_{\phi}\dot{A}^2&=0,\label{eomphi}\\
    \partial_t\left(af^2\dot{A}\right)+am_A^2A&=0,\label{eomA}
\end{align}
where $f_{\phi}\equiv\partial_{\phi}f$. On the other hand, Einstein's equations in an FRW universe are
\begin{align}
    3M_{\text{pl}}^2H^2&=\rho_{\phi}+\rho_A,\label{eom00}\\
    M_{\text{pl}}^2\left(2\dot{H}+3H^2\right)&=-\left(p_{\phi}+p_A\right),\label{eomij}
\end{align}
where $\rho_{\phi}=\dot{\phi}^2/2+V$ and $p_{\phi}=\dot{\phi}^2/2-V$ denote the energy density and pressure of the inflaton, respectively. In this work, we consider light vector bosons satisfying $m_A/f\ll H$, so that the mass term of the vector field can be neglected during inflation. In this limit, Eq.~\eqref{eomA} can be integrated once to give
\begin{align}
    af^2\dot{A}=p_A,
\end{align}
where $p_A$ is an integration constant.

If we assume that the energy density of the vector fields is negligible for both the spacetime dynamics and the inflaton evolution, i.e., $3M_{\text{pl}}^2H^2\simeq V$ and $3H\dot{\phi}+V_{\phi}\simeq0$ during inflation, the scale factor can be expressed as
\begin{align}
    a\propto\exp\left(-\frac{1}{M_{\text{pl}}^2}\int d\phi{}\frac{V}{V_{\phi}}\right),
\end{align}
where we have used Eqs.~\eqref{eomphi} and \eqref{eom00} in the slow-roll approximation. It then follows that the kinetic function scales as $f\propto a^{-2c}$. For negative $c$, $f$ increases as the Universe expands. Consequently, when extrapolated back to the early stage of inflation, $f$ may become arbitrarily small, leading to the strong coupling problem \cite{Demozzi:2009fu}. Therefore, in this work, we restrict our attention to positive values of $c$. In this case, the kinetic energy of the vector field evolves as $\rho_{A,\text{kin}}\propto a^{4(c-1)}$. Therefore, we consider $c\geq1$ to be of interest. 

In this case, the vector fields grow during inflation and eventually back-react on the inflaton dynamics. Specifically, in Eq.~\eqref{eomphi}, when the third term, which is of order $\rho_A(f_{\phi}/f)$, becomes comparable to the second term, the inflationary dynamics rapidly converge to a new attractor, in which the energy density of the vector fields is stabilized at a constant value, $\rho_{A,\text{kin}}\sim\text{const}$ \cite{Watanabe:2009ct,Chen:2021nkf,Firouzjahi:2018wlp,Gorji:2020vnh}.

In this new attractor, the kinetic function scales as $f\propto a^{-2}$, and the inflaton velocity is reduced to
\begin{align}\label{dphi}
    \dot{\phi}\simeq-\frac{V_{\phi}}{3cH}.
\end{align}
Although the vector fields back-react on the inflaton dynamics, their energy density remains much smaller than the total energy density driving inflation.

We define the ratio of the kinetic energies of the vector fields and the inflaton as
\begin{align}\label{h}
    h\equiv\sqrt{\frac{\rho_{A,\text{kin}}}{3\rho_{\phi,\text{kin}}}},
\end{align}
where $\rho_{\phi,\text{kin}}\equiv\dot{\phi}^2/2$ is the kinetic energy of the inflaton. From Eqs.~\eqref{eomphi} and \eqref{dphi}, we obtain
\begin{align}
    h\simeq\sqrt{\frac{c-1}{2}},
\end{align}
which remains approximately constant on this attractor.

\begin{figure}[t]
\centering
\includegraphics[scale=0.65]{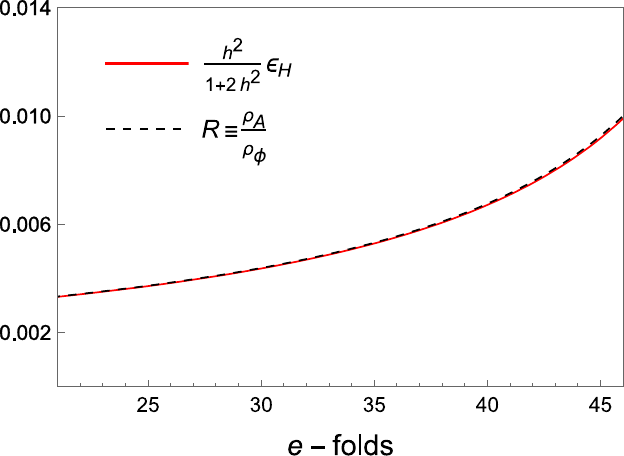}
\caption{\label{fig:R}Comparison of the evolution of $R$ and $h^2\epsilon_H/(1+2h^2)$. The parameters are chosen as $c=2$ and $V=m^2\phi^2/2$ with $m=10^{-5}M_{\rm pl}$. The initial conditions are $p_A=10^{45}$, $\phi_{\rm ini}=12M_{\rm pl}$, and $\dot{\phi}_{\rm ini}=0$.}
\end{figure}

We further define the ratio of the energy densities as $R\equiv{\rho_A}/{\rho_{\phi}}$. Using Eq.~\eqref{dphi}, we obtain
\begin{align}\label{Rdef}
    R\equiv\frac{\rho_A}{\rho_{\phi}}\simeq \frac{3h^2}{1+\frac{3}{\epsilon_V}(1+2h^2)^2},
\end{align}
where $\epsilon_V\equiv M_{\text{pl}}^2(V_{\phi}/V)^2/2$. On the other hand, from Eqs.~\eqref{eom00} and \eqref{eomphi}, we find
\[
\epsilon_H\equiv-\frac{\dot{H}}{H^2}\simeq\frac{\epsilon_V}{1+2h^2}.
\]
Therefore, in the slow-roll limit $\epsilon_H\ll1$, the ratio $R$ can be approximated as
\begin{align}
    R\simeq \frac{h^2}{1+2h^2}\epsilon_H=
    \begin{cases}
       h^2\epsilon_H,   & h\ll1, \\
       \epsilon_H/2,   & h\gg1.
    \end{cases}
\end{align}
We find that $\epsilon_H$ depends not on $c$ (or equivalently $h$ on this attractor), but on the initial value of $\rho_A$. As a result, the behavior of $R$ differs significantly between the small-$h$ and large-$h$ regimes. For $h\gg1$, $R$ depends only on $\epsilon_H$, whereas for $h\ll1$, it is proportional to both $h^2$ and $\epsilon_H$. The evolution of $R$ is illustrated in Fig.~\ref{fig:R}.

In the following, we do not impose any restriction on the magnitude of $h$ in the background evolution. However, this parameter plays a crucial role in determining the phenomenology of the vector fields. We therefore discuss the two regimes, $h\ll1$ (weak mixing) and $h\gg1$ (strong mixing), separately when studying the observational constraints.

\subsection{Dynamics after inflation}
Inflation ends when the inflaton rolls down to the minimum of its potential and begins to oscillate coherently. After inflation, the kinetic function approaches unity, $f(\phi)\to1$, so that the vector fields become decoupled from the inflaton. Reheating is completed when $H\sim\Gamma_{\phi}$, where $\Gamma_{\phi}$ is the decay rate of the inflaton.

After inflation, combining Eqs.~\eqref{eom00} and \eqref{eomij} yields
\begin{align}
    \dot{H}+\frac{3}{2}(1+w)H^2=0,
\end{align}
where $w\equiv p_{\text{tot}}/\rho_{\text{tot}}$ is the equation-of-state parameter of the total matter content. The corresponding solution is $H\propto a^{-3(1+w)/2}$. Defining the canonical field $A_c\equiv fA/a$, the equation of motion for the vector field after inflation becomes
\begin{align}
    \ddot{A}_c+3H\dot{A}_c+\left(m_A^2+\frac{1-3w}{2}H^2\right)A_c=0.
\end{align}

For $m_A\ll H$, the solution for the vector field is
\begin{align}
    A_c=c_1a^{(3w-1)/2}+c_2a^{-1}.
\end{align}
In both the radiation-dominated ($w=1/3$) and matter-dominated ($w=0$) eras, the energy density scales as $\rho_A\propto a^{-4}$. Therefore, the vector field behaves as a radiation component in this regime.

Since $\dot{H}<0$, the Hubble parameter eventually decreases to $H\sim m_A$, at which point the vector field begins to oscillate coherently. For $m_A\gg H$, the solution for $A_c$ can be expressed using the WKB approximation \cite{Ratra:1990me,Hwang:1996xd,Hwang:2009js},
\begin{align}
    A_c=\frac{d_1}{a^{3/2}}\sin(m_At)+\frac{d_2}{a^{3/2}}\cos(m_At).
\end{align}
The corresponding energy density is
\begin{align}
    \rho_A={}&\frac{3m_A^2}{2a^3}\bigg[(d_1^2+d_2^2)\bigg(1+\frac{H^2}{8m_A^2}\bigg)\nonumber\\
    &+\mathcal{C}\cos(2m_At)+\mathcal{S}\sin(2m_At)
     \bigg],
\end{align}
where $\mathcal{C}$ and $\mathcal{S}$ are functions of $d_1$, $d_2$, and $H/m_A$. Averaging the oscillation over a time scale of order $m_A^{-1}$, i.e. $\langle f(t)\rangle=(m_A/2\pi)\int_{0}^{2\pi/m_A}dt' f(t')$ and $\langle s(t)f(t)\rangle=s(t)\langle f(t)\rangle$ for rapid oscillation $f(t)$ and slow varying $s(t)$. The averaged energy density is then
\begin{align}
    \langle\rho_A\rangle =\frac{3m_A^2}{2a^3}\left[(d_1^2+d_2^2)+\mathcal{O}\left(\frac{H^2}{m_A^2}\right)\right].
\end{align}
Therefore, in the deep coherent oscillation regime, the vector field behaves as non-relativistic matter with $\langle\rho_A\rangle\propto a^{-3}$, and can thus serve as a viable dark matter candidate.

\section{Relic abundance of DM}\label{section3}
After inflation ends, the universe begins reheating. We assume that reheating ends at time $a_{\rm reh}$, which is after the end of inflation at $a_{\rm e}$. During this period, the inflaton coherently oscillates and behaves as a non-relativistic matter. Due to its coupling to the vector field, this process may generate additional particles in the vector field \cite{Kofman:1997yn,Braden:2010wd}. However, we assume that at $a_{\rm e}$, we have $f_{\rm e} = 1$ immediately, so the energy density of the vector field is given by its value at the end of inflation. 

Since the energy density of the vector field at the end of inflation is negligible compared with the inflationary energy, we consider that the energy of the universe during reheating is essentially provided by $\rho_{\phi}$. After reheating ends, this energy is instantaneously converted into radiation energy, which then characterizes the energy of the universe in the subsequent radiation-dominated era.

Keeping these in mind, we consider two cases: (1) the vector field oscillates after reheating ends $\Gamma_{\phi}>m_A$; (2) it oscillates before reheating ends $\Gamma_{\phi}<m_A$. We compute the ratio of the energy density of vector boson DM to the entropy density today \cite{Nakayama:2020rka}
\begin{align}
\frac{\rho_{A,0}}{s_0}\simeq4.4\times10^{-10}\text{GeV}.
\end{align}
Since $\rho_A\propto a^{-3}$ after $a_{\rm NR}$, where $a_{\rm NR}$ is the time when vector fields become non-relativistic, and $s\propto a^{-3}$, we have $\rho_{A}/s=\rm const$ after $a_{\rm NR}$. So we can instead compute $\rho_{A,\rm NR}/s_{\rm NR}$ in the following.

\subsection{$\Gamma_{\phi}>m_A$}
If the vector field begins to oscillate after reheating is completed, then at $a_{\rm NR}$ the universe is already in a radiation-dominated era, where the radiation energy originates from the decay of the inflaton $\phi$. Therefore $\rho_{\phi}\propto a^{-4}$ after $a_{\rm NR}$. Then,
\begin{align}
    \frac{\rho_{A,\rm NR}}{s_{\rm NR}}=\frac{\rho_{A,\rm reh}}{\rho_{\phi,\rm reh}}\frac{\rho_{\phi,\rm NR}}{s_{\rm NR}}=\frac{3R}{4}\frac{a_{\rm e}}{a_{\rm reh}}T_{\rm NR},
\end{align}
where we have used $\rho=(\pi^2/30)g_*T^4$ for radiation and $s=(\pi^2/45)g_*T^3$. The temperature  $T_{\rm NR}$ is given by
\begin{align}
    T_{\rm NR}=\left(\frac{90}{\pi^2g_*}\right)^{1/4}\sqrt{m_AM_{\text{pl}}}.
\end{align}
Also Using the Friedmann equation during the reheating era, $a_{\rm e}/a_{\rm reh}=(\Gamma_{\phi}/H_{\rm inf})^{2/3}$, where $H_{\rm inf}$ denotes the Hubble parameter during inflation. Then we finally get
\begin{align}\label{DMrelic1}
    \frac{\rho_{A,0}}{s_{0}}={}&4.4\rm GeV\nonumber\\
    &\times R\left(\frac{M_{\text{pl}}}{H_{\rm inf}}\right)^{2/3}\left(\frac{T_{\rm reh}}{10^{12}\rm GeV}\right)^{4/3}\left(\frac{m_A}{1\rm GeV}\right)^{1/2}.
\end{align}
Here, $T_{\rm reh}$ is the reheating temperature. $R$ is the energy-density ratio defined in Eq.~\eqref{Rdef}.

\subsection{$\Gamma_{\phi}<m_A$}
In this case, the vector field begins oscillate after reheating. We have assumed instantaneous reheating so $\rho_{\phi}\propto a^{-4}$ right after $a_{\rm reh}$. Then
\begin{align}
    \frac{\rho_{A,\rm NR}}{s_{\rm NR}}=\frac{\rho_{A,\rm NR}}{\rho_{\phi,\rm NR}}\frac{\rho_{\phi,\rm reh}}{s_{\rm reh}}=\frac{3R}{4}\frac{a_{\rm e}}{a_{\rm NR}}T_{\rm reh}.
\end{align}
By using $a_{\rm e}/a_{\rm NR}=(m_A/H_{\rm inf})^{2/3}$, we get
\begin{align}\label{DMrelic2}
    \frac{\rho_{A,0}}{s_{0}}={}&0.42\rm GeV\nonumber\\
    &\times R\left(\frac{M_{\text{pl}}}{H_{\rm inf}}\right)^{2/3}\left(\frac{T_{\rm reh}}{10^{12}\rm GeV}\right)\left(\frac{m_A}{1\rm GeV}\right)^{2/3}.
\end{align}

We have shown that, in both cases, the vector bosons can account for the observed dark matter abundance over a wide range of parameter space. We do not specify $H_{\rm inf}$ here because it depends on the background evolution discussed in Sec.~\ref{Dinf}. Instead, we treat it as a free parameter when deriving the observational constraints in the next section.

\section{Constrains from observations}\label{section4}
We now consider the cosmological constraints on this model. Since the vector-field misalignment is produced during inflationary period, we will make use of the power spectrum and non-Gaussianities of the curvature fluctuation, and the CMB constraint on isocurvature fluctuation from large-scale CMB observations to discuss our constraints. Before doing this, we should first mention that the time of back-reaction of vector fields to the inflaton $N_{\rm back}$ depends on the initial conditions and model parameters (e.g. $c$ in our model). We assume $N_{\rm back}>60$, such that when the modes relevant to CMB observations exit the horizon, inflation has already entered the new attractor solution described by Eq.~\eqref{dphi}. Second, we consider only the leading order of the slow-roll approximation in our constraints. For example, without vector fields, the power spectrum is scale invariant.

In the above discussion of background evolution, we have already separated the two cases, $h\ll1$ and $h\gg1$. Distinguishing between these two regimes is phenomenologically very important. In our previous work \cite{Chen:2022ccf,Chen:2023bcz,Chen:2026rwr,Chen:2025qyv}, we have found that the two cases contribute entirely different types of modes to the CMB: for $h\ll1$, these modes are of the accumulative modes, whereas for $h\gg1$, they are constant modes. Moreover, in the $h\gg1$ case, these modes experience exponential growth before horizon crossing. Therefore, it is necessary to discuss the constraints for these two cases separately.

From the last section, we need to know $R$, which is related to $h$ and $\epsilon_H$, to get a correct relic of DM. We found that $\epsilon_H$ is independent of $h$, but on the initial conditions $\rho_{A,\rm ini}$, while $H_{\rm inf}$ depends both on $h$ and the $\rho_{A,\rm ini}$. So we treat $R$ as a function of two free parameters $h$ and $H_{\rm inf}$ in our constraints. We also fixed $T_{\rm reh}=10^{12}\rm GeV$ in the following discussion.

\subsection{Curvature fluctuation}
For a nonzero vector-field energy density $\rho_A$ with kinetic coupling during inflation, the fluctuations of inflaton $\delta\phi$ and vector fields $\delta A$ are generally coupled to each other, with the mixing strength controlled by $h$ (see appendix \ref{app:longitudinal}). In this case, both the inflaton and the vector-field fluctuations contribute to the curvature fluctuation, with their relative contributions determined by $h$ \cite{Chen:2022ccf,Chen:2023bcz,Chen:2026rwr}:
\begin{align}
    \mathcal{R}\simeq\frac{1}{\sqrt{2\epsilon_{H}}M_{\text{pl}}}\frac{-\delta\phi+\sqrt{2}h\mathcal{A}_T}{\sqrt{1+2h^2}},
\end{align}
where $\mathcal{A}_T\equiv\sqrt{2}f\delta A/a$ and $\delta A$ denotes the transverse mode of vector fields, and corresponds to the isotropized fluctuation of $A$ (see appendix \ref{app:longitudinal}). 

\begin{figure}[t]
\centering
 \begin{minipage}{1\linewidth}
\centering
\includegraphics[scale=0.86]{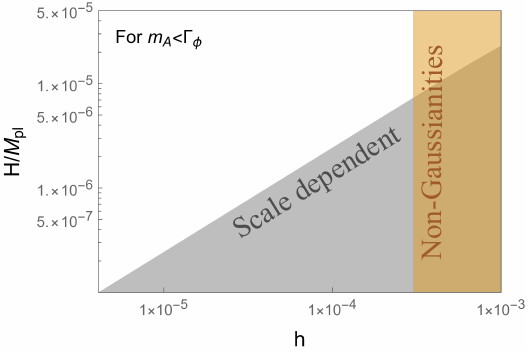}
\caption{\label{fig:C_small_h}Constraints on the curvature fluctuation for $h\ll 1$. The gray region is excluded by the scale-invarince of the power spectrum. The orange region is excluded by the amplitudes of bispectrum.}
\end{minipage}\\
\vspace{0.5cm} 
\begin{minipage}{1\linewidth}
    \centering
\includegraphics[scale=0.86]{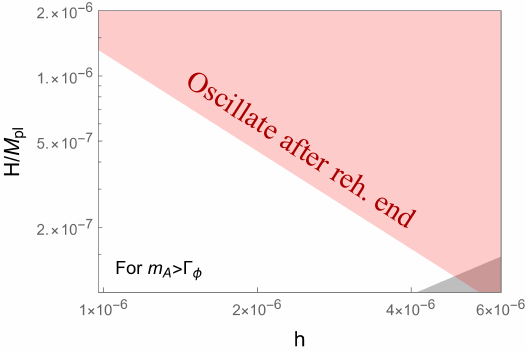}
\caption{\label{fig:C_small_h2}Constraints on the curvature fluctuation for $h\ll 1$ and fixing $T_{\rm reh}=10^{12}\rm GeV$. The red region is excluded by the $m_A>\Gamma_{\phi}$. The gray region is excluded by the scale-invarince of the power spectrum shown in fig. \ref{fig:C_small_h}.}
\end{minipage}
\end{figure}

\subsubsection{$h\ll1$}
In this regime, the mixing between the inflaton and vector-field fluctuations can be treated perturbatively. The power spectrum of the curvature perturbation can therefore be computed using the in-in formalism by expanding around the $h=0$ limit, yielding \cite{Chen:2023bcz,Chen:2026rwr}
\begin{align}
    \mathcal{P}_{\mathcal{R}}=\frac{H_{\rm inf}^2}{8\pi^2\epsilon_HM_{\rm pl}^2}\left(1+32h^2N_k^2\right),
\end{align}
where $N_k=-\ln(-k\tau_{\rm e})$, with $\tau_{\rm e}$ denoting the conformal time at the end of inflation. The quantity $N_k$ measures the number of e-folds between horizon crossing of the mode $k$ and the end of inflation. In other words, this correction arises from a cumulative effect: the curvature fluctuation is sourced by vector fluctuation on superhorizon scales and still evolves until the end of inflation. 

Although the isotropized vector-field background preserves the statistical isotropy of the power spectrum, the scale dependence of the spectrum is still modified. The corresponding correction to the spectral index is
\begin{align}
    \Delta n_s=\frac{d\ln(\Delta\mathcal{P}_{\mathcal{R}})}{d\ln k}\bigg|_{k=k_*}=64h^2N_k.
\end{align}

Requiring this correction to be no larger than the slow-roll contribution, namely $\Delta n_s\lesssim\epsilon_H$, gives $h<\sqrt{\epsilon_H/}60$ for $N_k\sim60$. The amplitude of the curvature perturbation has been measured by Planck 2018 to be $\mathcal{P}_{\mathcal{R}}\simeq2.1\times10^{-9}$ at the pivot scale $k=k_*$ \cite{Planck:2018jri}. The resulting constraint from the spectral index is shown in Fig.~\ref{fig:C_small_h}.

At this point,  we can express $\epsilon_H$ in terms of $h$ and $H_{\rm inf}$. The ratio $R$ in Eqs.~\eqref{DMrelic1} and \eqref{DMrelic2} can then be written as
\begin{align}
    R\simeq\frac{h^2(1+32h^2N_k^2)}{8\pi^2\mathcal{P}_{\mathcal{R}}}\left(\frac{H_{\rm inf}}{M_{\rm pl}}\right)^2.
\end{align}
Then the mass of vector boson is given by
\begin{widetext}
    \begin{align}
    m_A\simeq
    \begin{cases}
       6.0\times10^{-21}\rm GeV\times\left(\frac{H_{\rm inf}}{10^{-5}M_{\rm pl}}\right)^{-8/3}\left(\frac{T_{\rm reh}}{10^{12}\rm GeV}\right)^{-8/3}h^{-4},   & \Gamma_{\phi}>m_A, \\
       2.3\times10^{-14}\rm GeV\times\left(\frac{H_{\rm inf}}{10^{-5}M_{\rm pl}}\right)^{-2}\left(\frac{T_{\rm reh}}{10^{12}\rm GeV}\right)^{-3/2}h^{-3},   & \Gamma_{\phi}<m_A.
    \end{cases}
\end{align}
\end{widetext}
In both cases, $\Gamma_{\phi}>m_A$ and $\Gamma_{\phi}<m_A$, the vector-boson mass increases as $h$, $H_{\rm inf}$, or $T_{\rm reh}$ decreases. The upper bound on $H_{\rm inf}$ is constrained by the tensor-to-scalar ratio $r$ \cite{BICEP:2021xfz},
\begin{align}
    H_{\rm inf}\lesssim 10^{-5}M_{\rm pl}.\ \ \ \ \ (r<0.036)
\end{align}
Fixing $T_{\rm reh}=10^{12}\rm GeV$ and considering the regime $h<1$, we find that for $\Gamma_{\phi}>m_A$, the vector-boson mass ranges from approximately $10^{-21}\rm GeV$ up to $\Gamma_{\phi}\sim10^5\rm GeV$. As we discuss below, however, the non-Gaussianity constraints exclude such ultralight vector bosons. In contrast, for $\Gamma_{\phi}<m_A$, the lower bound on the vector-boson mass is shown in Fig.~\ref{fig:C_small_h2}, while the upper bound is determined by the limit on $H_{\rm inf}$. Therefore, in this scenario the vector dark matter is generally heavier than in the $\Gamma_{\phi}>m_A$ case.

\begin{figure}[t]
\centering
\includegraphics[scale=0.78]{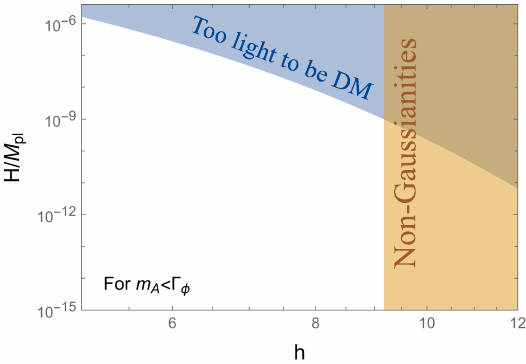}
\caption{\label{fig:C_large_h}Constraints on the curvature fluctuation for $h\ll 1$ and $T_{\rm reh}=10^{12}\rm GeV$. The blue region is excluded by $m_A>H_{\rm eq}$. The orange region is excluded by the amplitudes of bispectrum.}
\end{figure}
\subsubsection{$h\gg1$}
In this regime, the cumulative mode decays away and is replaced by the constant mode. In other words, the curvature fluctuation does not evolve on superhorizon at leading order of the slow-roll approximation. Moreover, the curvature mode experiences a transient growth before crossing horizon. The scale-invariant power spectrum for $h\gtrsim5$ can be computed as \cite{Chen:2022ccf,Chen:2023bcz}
\begin{align}\label{PR_large_h}
    \mathcal{P}_{\mathcal{R}}=\frac{H_{\rm inf}^2}{8\pi^2\epsilon_HM_{\rm pl}^2}e^{2.37h}.
\end{align}
which does not depend on the number of e-folds, but an exponential factor $e^h$. That is, at this point, we are free from scale-dependent constraints at leading order of slow-roll. 

In this case, the ratio $R$ can be represented by
\begin{align}
    R\simeq\frac{1}{16\pi^2\mathcal{P}_{\mathcal{R}}}\left(\frac{H_{\rm inf}}{M_{\rm pl}}\right)^2e^{2.37h}.
\end{align}
Then we can use this to estimate the mass of vector boson. For $h\gtrsim5$, the mass of the vector boson is
\begin{widetext}
    \begin{align}
    {m_A}\simeq
    \begin{cases}
        2.4\times10^{-20}\rm GeV\times\left(\frac{H_{\rm inf}}{10^{-5}M_{\rm pl}}\right)^{-8/3}\left(\frac{T_{\rm reh}}{10^{12}\rm GeV}\right)^{-8/3}e^{-4.74h},  &\Gamma_{\phi}>m_A\\
        6.5\times10^{-14}\rm GeV\times\left(\frac{H_{\rm inf}}{10^{-5}M_{\rm pl}}\right)^{-2}\left(\frac{T_{\rm reh}}{10^{12}\rm GeV}\right)^{-3/2}e^{-3.56h},   &\Gamma_{\phi}<m_A
    \end{cases}
\end{align}
\end{widetext}
For large-$h$, the mass can be ultra-light due to the exponential suppression. However, for $\Gamma_{\phi}>m_A$, the vector boson can serve as dark matter only if it becomes non-relativistic before matter-radiation equality, i.e. $m_A>H_{\rm eq}$, where $H_{\rm eq}\simeq1.5\times10^{-28}\rm GeV$ is the Hubble parameter when the energy density of radiation and matter is equal. This provides the constraint parameter space shown in fig. \ref{fig:C_large_h}. On the other hand, for $\Gamma_{\phi}<m_A$, we find that the required inflationary Hubble scale is too small ($H_{\rm inf}<10^{-19}M_{\rm pl}$) to admit any viable parameter space. Therefore, the $\Gamma_{\phi}<m_A$ scenario is excluded in the strong-mixing regime, $h\gg1$..

\subsection{Non-Gaussianities}
Another important observational probe of the curvature fluctuation is the non-Gaussianities. As a result of the interactions between the inflaton and the vector fields, the curvature fluctuation can acquire non-Gaussianity. We also divide the discussion into the two regimes of $h$, because the contributions to the non-Gaussianities arise from different modes. 

\subsubsection{$h\ll1$}
In this regime, we can compute the non-Gaussianities perturbatively by treating mixing terms as small corrections. The dominant contribution of the bispectrum of the curvature fluctuation is given by \cite{Chen:2026rwr}
\begin{align}
     B_{\mathcal{R}}(k_1,k_2,k_3)\simeq{}&256h^2N_K^3(2\pi^2)^2\mathcal{P}_{\mathcal{R}}^2\nonumber\\
     &\times\frac{k_1^3+k_2^3+k_3^3}{k_1^3k_2^3k_3^3}+\mathcal{O}(N_K^2),
\end{align}
where $N_K\equiv-\log(-K\tau_{\rm e})$ and $K\equiv(k_1+k_2+k_3)/3$. This is a local type non-Gaussianities, and its amplitude is given by
\begin{align}
f_{\rm NL}^{\rm local}=512h^2N_K^3.
\end{align}
Similar to the power spectrum, the non-Gaussianities also depend on the e-folds number after exiting the horizon due to the evolution of superhorizon $\mathcal{R}$. For $N_K\sim60$, the observations $|f_{\rm NL}^{\rm local}|\lesssim 10$ impose a stringent constraint $h<3\times10^{-4}$, as shown in fig. \ref{fig:C_small_h}.

We have mentioned that $m_A$ decreases with increasing $h$ and $H_{\rm inf}$. For $\Gamma_{\phi}>m_{A}$, we found that in the allowed parameter space, the "lower bound of the vector boson mass becomes $10^{-6}\rm GeV$.

\subsubsection{$h\gg1$}

In this regime, the constant mode gives rise to a more complicated shape of non-Gaussianities. Unlike the power spectrum, which is exponentially enhanced by $h$, the non-Gaussianities remain relatively mild for large-$h$ \cite{Chen:2026rwr}. Here, we constrain the parameters from the local shape and the equilateral shape. The corresponding bistpecrum are given by \cite{Chen:2026rwr}
\begin{align}
    B^{\rm local}_{\mathcal{R}}\simeq{}&(2\pi^2)^2\left(14-4h-\frac{12}{h}\right)\frac{\mathcal{P}_{\mathcal{R}}^2}{(k_Sk_L)^3},\\
    B^{\rm equil}_{\mathcal{R}}\simeq{}&(2\pi^2)^2\frac{178}{9}\frac{\mathcal{P}_{\mathcal{R}}^2}{k^6},
\end{align}
where in the local limit $k_1=k_2=k_S$ and $k_3=k_L\ll k_S$, while in the equilateral limit we take $k_1=k_2=k_3=k$. the corresponding amplitudes are then given by
\begin{align}
    f_{\rm NL}^{\rm local}=\frac{5}{6}\left(7-2h-\frac{6}{h}\right),\ \ \ \ \ \ 
    f_{\rm NL}^{\rm equil}=\frac{445}{54}.
\end{align}

The local non-Gaussianity is depends only polynomially on $h$, while the equilateral one is a constant. These results are computed under an effective field theory description of the original system, which is valid for $h>5$. In this regime, we find $f_{\rm NL}^{\rm local}<0$. From the constraint $|f_{\rm NL}^{\rm local}|\lesssim 10$ we should have $h<9.2$. This constraint is also shown in fig. \ref{fig:C_large_h}. On the other hand, $f_{\rm NL}^{\rm equil}$ is well within the current observational bounds.

\subsection{Isocurvature fluctuation}
Now we consider the constraints from isocurvature fluctuation. Similar to the case of multi-field inflation, one can define the entropy fluctuation that is ``orthogonal'' to the curvature fluctuation as 
\begin{align}
    \mathcal{S}\equiv -\frac{1}{\sqrt{2\epsilon_{H}}M_{\text{pl}}}\frac{\mathcal{A}_T+\sqrt{2}h\delta\phi}{\sqrt{1+2h^2}}.
\end{align}
In addition, there exists a magnetic fluctuation, denoted by $U$, which is decoupled from other fluctuations (see appendix \ref{app:longitudinal}). This is a massless field, so t is also expected to contribute to the isocurvature fluctuation. The power spectrum of fluctuation $U$ is the same for weak- and strong-mixing regimes, and it is given by
\begin{align}
    \mathcal{P}_{U}=\frac{H_{\rm inf}^2}{8\pi^2\epsilon_HM_{\rm pl}^2},
\end{align}
while the entropy fluctuation is quite different for these two cases, hence we need to discuss it separately. The power spectrum of the longitudinal mode is strongly blue-tilted, so can be ignored \cite{Nakayama:2019rhg}.

\subsubsection{$h\ll1$}
In this case, the computation is"similar to that of the curvature fluctuation, which treats vector fields as a small correction to the inflaton sector and then using the in-in formalism around the free theory. The power spectrum of the entropy perturbation is then given by \cite{Gorji:2020vnh}
\begin{align}
    \mathcal{P}_{\mathcal{S}}=\frac{H_{\rm inf}^2}{8\pi^2\epsilon_HM_{\rm pl}^2}\left(1-\frac{56}{3}h^2N_k^2\right).
\end{align}
The total isocurvature power spectrum is the sum of the entropy and the magnetic fluctuations $\mathcal{P}_{\mathcal{I}}=\mathcal{P}_{\mathcal{S}}+\mathcal{P}_{U}$. Moreover, the cross-correlation between curvature and isocurvature fluctuations vanishes, which means these two fluctuations are uncorrelated \cite{Gorji:2020vnh}, that is, $\mathcal{P}_{\mathcal{R}\mathcal{I}}=0$.

The uncorrelated isocurvature power spectrum is constrained by CMB observations as \cite{Planck:2018jri}
\begin{align}
    \beta\equiv\frac{\mathcal{P}_{\mathcal{I}}}{\mathcal{P}_{\mathcal{R}}+\mathcal{P}_{\mathcal{I}}}<0.038\ \ \ \ (\rm uncorrelated),
\end{align}
which gives $h>5.2\times10^{-3}$. This region is already excluded by the constraint of non-Gaussianities on curvature fluctuation shown in fig. \ref{fig:C_small_h}.

\begin{table*}[!t]
    \caption{Summary of the constraints and the mass of vector DM.}
    \label{tab:conclusion}
    \centering
    \renewcommand{\arraystretch}{1.5}
    \setlength{\tabcolsep}{9pt}
    \begin{tabular}{c|c|c|c|c|c}
        \hline
        \multicolumn{2}{c|}{} &
        \multicolumn{2}{c|}{Curvature} &
        \multirow{2}{*}{Isocurvature} &
        \multirow{2}{*}{Mass of vector DM}
        \\
        \cline{3-4}
        \multicolumn{2}{c|}{} &
        Power spectrum &
        Bispectrum &
        &
        \\
        \hline

        \multirow{2}{*}{$h \ll 1$}
        &
        $m_A < \Gamma_{\phi}$
        &
        Fig. \ref{fig:C_small_h}
        &
        \multirow{2}{*}{$h < 3 \times 10^{-4}$}
        &
        \multirow{2}{*}{$h > 5.2 \times 10^{-3}$}
        &
        \multirow{2}{*}{
            $\displaystyle
            \frac{m_A}{\mathrm{GeV}}
            >
            7.4 \times 10^{-7}
            \left(
                \frac{T_{\mathrm{reh}}}
                {10^{12}\,\mathrm{GeV}}
            \right)^{-8/3}$
        }
        \\
        \cline{2-3}
        &
        $m_A > \Gamma_{\phi}$
        &
        Fig. \ref{fig:C_small_h2}
        &
        &
        &
        \\
        \hline

        \multirow{2}{*}{$h \gg 1$}
        &
        $m_A < \Gamma_{\phi}$
        &
        Fig. \ref{fig:C_large_h}
        &
        \multirow{2}{*}{$h < 9.2$}
        &
        \multirow{2}{*}{$h > 45$}
        &
        \multirow{2}{*}{
            $\displaystyle
            \frac{m_A}{\mathrm{GeV}}
            >
            1.5 \times 10^{-28}$
        }
        \\
        \cline{2-3}
        &
        $m_A > \Gamma_{\phi}$
        &
        None
        &
        &
        &
        \\
        \hline
    \end{tabular}
\end{table*}

\subsubsection{$h\gg1$}
For large-$h$, the cumulative mode decays away on superhorizon scales. Instead, the mode $\mathcal{S}$ is also a constant mode on super-horizon with amplitude \cite{Chen:2023bcz}
\begin{align}\label{SR}
    \mathcal{S}\simeq \frac{16\sqrt{2}h^2H^2}{m_s^2(1+2h^2)}\mathcal{R},
\end{align}
where the mass of the entropy fluctuation $m_s^2=-8H^2h^2(3+2h^2)/(1+2h^2)<0$. For large $h$, we have the power spectrum of the entropy fluctuation
\begin{align}
    \mathcal{P}_{\mathcal{S}}\simeq\frac{2}{h^2}\mathcal{P}_{\mathcal{R}},
\end{align}
where the power spectrum $\mathcal{P}_{\mathcal{R}}$ is given by Eq.~\eqref{PR_large_h}. Since the entropy fluctuation is exponentially enhanced, the contribution from the magnetic mode $U$ is negligible in the isocurvature fluctuation, so that $\mathcal{P}_{\mathcal{I}}\simeq\mathcal{P}_{\mathcal{S}}$. We also note from this relation Eq.~\eqref{SR} that the curvature and the entropy fluctuations are anti-correlated. Hence, the constraint on the isocurvature fluctuation \cite{Planck:2018jri}
\begin{align}
    \beta\equiv\frac{\mathcal{P}_{\mathcal{I}}}{\mathcal{P}_{\mathcal{R}}+\mathcal{P}_{\mathcal{I}}}<10^{-4}\ \ \ \ (\text{anti-correlated})
\end{align}
implies $h>45$, which is incompatible with the constraint of non-Gaussianities shown in fig. \ref{fig:C_large_h}.

\section{Conclusions}\label{section5}
We have already discussed the possibility of coherently oscillating vector bosons as a candidate for DM. The mechanism is similar to the misalignment mechanism of axion DM production. We consider the isotropized vector-field configuration to preserve the constraints from the large-scale anisotropy. This can be realized by a number of vector fields, so that the field configuration becomes isotropic after averaging.

We discuss weak- and strong-mixing regimes, respectively, where $h$ is the mixing strength between perturbations defined by Eq.~\eqref{h}. Previous studies have revealed that the modes that contribute to the observations are quite different in these two cases. For small-$h$, the dominant contribution is the cumulative modes, so that the power spectrum of the curvature is still constrained by the scale-invariant. On the other hand, for large-$h$, the dominant modes are constant modes, and it is free from scale-dependent constraints from the power spectrum of the curvature fluctuation. However, we should put a bound from $H_{\rm eq}$ for the vector bosons to behave as DM.

We also take into account the constraints of curvature non-Gaussianities and isocurvature fluctuation for these two cases. Unfortunately, we found that in both cases, non-Guassianities and isocurvature fluctuation are incompatible with each other. The results of the constraints are summarized in the table \ref{tab:conclusion}.

So for we have found that coherently oscillating vector bosons are unlikely to provide a viable model for DM candidates. Although we use isotropized vector configuration to avoid anisotropy, we still meet the issue of overproduction of isocurvature fluctuation once the non-Gaussianities constraints are imposed. A possible loophole for this issue is the vector curvaton, which means a fraction of vector fields may contribute to the curvature fluctuation rather than DM. We leave an investigation of this possibility for future work.

\begin{acknowledgments}
I would like to thank Yong-Chao Li for teaching me in the art of pour‑over coffee during the final stage of this work, which made the completion of this work possible.
\end{acknowledgments}

\appendix
\section{Cosmological fluctuations}\label{app:longitudinal}
We can decompose the scalar parts of fluctuations as follows \cite{Maleknejad:2011sq,Maleknejad:2011jw},
\begin{align}
\phi={}&\phi(t)+\delta\phi,\ \ \ \ A^a_{\ 0}=\partial_aY,\nonumber\\
A^a_{\ i}={}&\left(A+\delta A\right)\delta_{ai}+\epsilon_{iab}\partial_bU+\partial_i\partial_aM.
\end{align}
We define the ``longitudinal'' mode by projecting gauge fields on the direction of wavenumber 
\begin{equation}
    A_L\equiv-\frac{k_ak_i}{k^2}A^a_{\ i}=k^2M-\delta A.
\end{equation}
After fixing the spatially-flat gauge and eliminating the non-dynamical mode $Y$, the quadratic action can be written as
\begin{align}
\label{QC}
\mathcal{L}^{(2)}_{\text{tot}}=&\frac{a^3}{2}\Bigg\{\delta\dot{\phi}^2+\bigg(-\frac{k^2}{a^2}+8h^2H^2\bigg)\delta\phi^2+\dot{\mathcal{A}_T}^2\nonumber\\
&-\left(1+\lambda\right)\frac{k^2}{a^2}\mathcal{A}_T^2+8\sqrt{2}hH\delta\phi \dot{\mathcal{A}_T}
    +24\sqrt{2}hH^2\delta\phi\mathcal{A}_T\nonumber\\
&+ \dot{\mathcal{A}}_L^2-\bigg[\left(1+\lambda\right)\frac{k^2}{a^2}-\frac{1}{2}\frac{\ddot{\lambda}}{(1+\lambda)\lambda}+\frac{1}{4}\frac{\dot{\lambda}^2}{(1+\lambda)^2\lambda^2}\nonumber\\
&+\frac{\dot{\lambda}^2}{(1+\lambda)^2\lambda}+\frac{3}{2}\frac{\dot{\lambda}}{(1+\lambda)\lambda}H\bigg] \mathcal{A}_L^2+\dot{\mathcal{U}}^2-\frac{k^2}{a^2}\mathcal{U}^2\Bigg\},
\end{align}
where we have defined $\mathcal{A}_L\equiv f\sqrt{\lambda/(1+\lambda)}A_L/a$, $\mathcal{A}_T\equiv\sqrt{2}f\delta A/a$, $\mathcal{U}\equiv\sqrt{2}f\partial_iU/a$, and 
\begin{align}
    \lambda\equiv \frac{a^2m_A^2}{f^2k^2}.
\end{align}
In the Lagrangian of $\delta\phi$ and $\mathcal{A}_T$, the term $\lambda k^2/a^2=m_A^2/f^2\ll H_{\rm inf}^2$. Hence, we can ignore this term when we consider the dynamics of $\delta\phi$ and $\mathcal{A}_T$ during inflation.

Then after defining the canonical curvature $\mathcal{R}_c\equiv\sqrt{2\epsilon_H}M_{\rm pl}\mathcal{R}$ and entropy $\mathcal{S}_c\equiv\sqrt{2\epsilon_H}M_{\rm pl}\mathcal{S}$ fluctuations, the quadratic Lagrangians of $\delta\phi$ and $\mathcal{A}_T$ can be expressed in terms of the curvature, entropy fluctuations as
\begin{align}\label{L2}
    \mathcal{L}^{(2)}_{\mathcal{R}\mathcal{S}}={}&\frac{a^3}{2}\bigg[\dot{\mathcal{R}}_c^2-\frac{k^2}{a^2} \mathcal{R}_c^2 -m_{R}^2\mathcal{R}_c^2+\dot{\mathcal{S}}_c^2-\frac{k^2}{a^2}\mathcal{S}_c^2-m_s^2\mathcal{S}_c^2\nonumber\\
    &-2\sqrt{2}hH\left(4\dot{\mathcal{R}}_c
    +\frac{16h^2}{1+2h^2}H\mathcal{R}_c\right)\mathcal{S}_c\bigg],
\end{align}
where $m_R^2=16h^2H^2/(1+2h^2)$ and $m_s=-8h^2(3+2h^2)H^2/(1+2h^2)$. Also, the quadratic Lagrangian of magnetic fluctuation $U$ is given by
\begin{align}
    \mathcal{L}_U^{(2)}=\frac{a^3}{2}\left(\dot{\mathcal{U}}^2-\frac{k^2}{a^2}\mathcal{U}^2\right).
\end{align}

On the other hand, in the ``anisotropic'' attractor we have $f=(a_e/a)^2$ so we can obtain $\dot{\lambda}=6H\lambda$ and $\ddot{\lambda}=36H^2\lambda$. Then the quadratic Lagrangian of the longitudinal mode is given by
\begin{align}
    \mathcal{L}_{\mathcal{A}_L}^{(2)}=\frac{a^3}{2}\left[\dot{\mathcal{A}}_L^2-\frac{k^2(1+\lambda)}{a^2}\mathcal{A}_L^2-\frac{27H^2\lambda}{(1+\lambda)^2}\mathcal{A}_L^2\right].
\end{align}
It is shown that the power spectrum of longitudinal modes is strongly blue tilted \cite{Nakayama:2019rhg}, so can be ignored compared with the transverse one at the present cosmological scales.

\nocite{*}

\bibliography{Iso_CVODM_ref}

\end{document}